\documentclass[12pt]{iopart}

\usepackage{graphicx}
\usepackage{iopams}
\usepackage{cite}
\usepackage{color}
\usepackage[overlay,absolute]{textpos}
\definecolor{mygray}{gray}{0.5}

\newcommand{\thetaf}[4]{%
\theta\!\left[ #1 \atop #2 \right]\!\left(#3,#4\right)}

\begin{document}

\begin{textblock*}{\textwidth}[0,0](25.5mm,32mm)
\footnotesize\noindent
\begin{minipage}{\textwidth}
\center
\textcolor{mygray}{Journal link: http://dx.doi.org/10.1088/1742-5468/2014/04/P04007}
\end{minipage}
\end{textblock*}

\begin{textblock*}{\textwidth}[0,0](25.5mm,268mm)
\footnotesize\noindent
\begin{minipage}{\textwidth}
\textcolor{mygray}{Journal ref: A. E. B. Nielsen and G. Sierra, J. Stat.\ Mech.\ \textbf{2014}, P04007 (2014).}\\
\textcolor{mygray}{\copyright{ }2014 IOP Publishing Ltd and SISSA Medialab srl.}
\end{minipage}
\end{textblock*}

\title[Bosonic FQH states on the torus from CFT]{Bosonic fractional quantum Hall states on the torus from conformal field theory}
\author{Anne E B Nielsen$^1$ and Germ\'an Sierra$^2$}
\address{$^1$ Max-Planck-Institut f{\"u}r Quantenoptik,
Hans-Kopfermann-Stra{\ss}e 1, D-85748 Garching, Germany}
\address{$^2$ Instituto de F\'isica Te\'orica, UAM-CSIC, Madrid, Spain}

\begin{abstract}
The Kalmeyer-Laughlin state, which is a lattice version of the bosonic Laughlin state at filling factor one half, has attracted much attention due to its topological and chiral spin liquid properties. Here we show that the Kalmeyer-Laughlin state on the torus can be expressed in terms of a correlator of conformal fields from the $SU(2)_1$ Wess-Zumino-Witten model. This reveals an interesting underlying mathematical structure and provides a natural way to generalize the Kalmeyer-Laughlin state to arbitrary lattices on the torus. We find that the many-body Chern number of the states is unity for more different lattices, which suggests that the topological properties of the states are preserved when the lattice is changed. Finally, we analyze the symmetry properties of the states on square lattices.
\end{abstract}

\section{Introduction}\label{introduction}

In \cite{MR}, Moore and Read showed that a number of fractional quantum Hall (FQH) states can be expressed as certain correlators of conformal fields. More recently it has been shown that such a connection also holds for FQH states in lattice systems \cite{CS,nsc2} and for the interpolation between lattice and continuum systems \cite{tu2}. The construction is interesting because it shows that the states can be seen as special cases of a more general mathematical framework. It also means that the states fulfil certain mathematical relations inherited from conformal field theory (CFT), and these are useful for deriving various properties of the states. Examples include the possibility to derive parent Hamiltonians, to study quasi particle or edge excitations, to investigate entanglement spectra and to derive correlation functions \cite{MR,CS,nsc2,tu2,wen1,wen2,nsc1,tu1}.

We shall here focus on a lattice version of the bosonic Laughlin state at filling factor $1/2$, which is known as the Kalmeyer-Laughlin (KL) state \cite{KL87,KL89}. On the Riemann sphere, i.e., the complex plane combined with the point at infinity, the KL state takes the form
\begin{equation}\label{KLsphere}
\psi_\mathrm{KL}(Z_1,Z_2,\ldots,Z_M)=\prod_iG(Z_i) \prod_{i<j}(Z_i-Z_j)^2\prod_i\rme^{-\frac{1}{4}|Z_i|^2}
\quad (\mathrm{plane}).
\end{equation}
This wavefunction describes a state of $M$ bosonic particles on a two-dimensional square or triangular lattice, and $Z_i$ is the position of the $i$th particle written as a complex number. The factor $\prod_iG(Z_i)$ is a gauge factor that is included to ensure that the state is a singlet, $G(Z_i)\in\{-1,+1\}$ for all $i$ and the rest of the expression on the right hand side of \eref{KLsphere} is the bosonic Laughlin state at filling factor $1/2$, except that the possible values of the positions $Z_i$ are restricted to the considered lattice. The KL state has also been studied on a square lattice on the torus \cite{L89}, and we provide an expression for this state in section \ref{Sec:KL}. The KL state and closely related states have been analyzed in several papers, see e.g.\ \cite{L89,wen,zou,balatsky,wen91,barkeshli,nsc3}. These analyses show that the KL state has the same topological properties as the bosonic Laughlin state at half filling in the continuum. The investigated properties include, e.g., correlation functions, the fractional statistics of quasiparticle excitations, the chiral edge states, and topological entanglement entropy. Parent Hamiltonians have been obtained in \cite{nsc2,nsc3,schroeter,thomale,kapit,bauer}.

The huge interest in the KL state is due to its topological and chiral spin liquid properties combined with a relatively simple analytical wavefunction defined on a lattice. To characterize the properties of topological quantum states a number of different measures have been proposed, but more of these requires the state under investigation to be defined with a particular set of boundary conditions, e.g.\ periodic boundary conditions that correspond to defining the state on a torus. It is therefore very relevant to study different geometries.

In the present paper, we demonstrate that the KL states on the torus can be expressed as chiral correlators of fields from the $SU(2)_1$ Wess-Zumino-Witten (WZW) model. It has previously been shown \cite{balatsky,nsc2,tu2} that this construction gives a KL-like state when considered on the Riemann sphere, and the present study extends this result to the torus geometry, where the relation to the KL state turns out to be exact. The CFT correlators can be defined for arbitrary lattices in 2D and are ensured to be singlets by construction. The CFT states thus provide a generalization of the KL states on the torus to arbitrary lattices, and by investigating the topological properties of the states on different lattices, we provide evidence that the topology remains the same at least for a broad class of lattices. For the case of square lattices with $L_x \times L_y$ spins, we also find linear combinations of the CFT states that are eigenstates of various symmetry operators. In particular, this allows us in certain cases to identify linear combinations of the states that are guaranteed to be orthogonal.

The paper is structured as follows. In section \ref{Sec:wfcft}, we explain how wavefunctions can be constructed from conformal fields. We also introduce the CFT states that we study in this work and provide analytical expressions for them on the Riemann sphere and torus geometries. In section \ref{Sec:cblock}, we derive the analytical expressions for the CFT states on the torus. In section \ref{Sec:KL}, we show that the CFT states are proportional to the KL states on the torus when defined on a square lattice. In section \ref{Sec:trans}, we study the symmetry properties of the CFT states for the case of a square lattice. In particular, we find linear combinations of the two states that are eigenstates of different transformation operators and their eigenvalues. In section \ref{Sec:chern}, we discuss the topological properties of the states, and we find among other things that the many-body Chern number of the states is unity for several different lattices. Section \ref{Sec:conclusion} concludes the paper, \ref{Sec:AppA} summarizes the definition and some properties of Riemann theta functions and \ref{Sec:AppB} provides an alternative derivation of the singlet property of the states for four spins. To provide readers who are more interested in the concepts than the derivations with a faster route through the paper, we start sections \ref{Sec:cblock}, \ref{Sec:KL} and \ref{Sec:trans} with a summary of the results derived in the section.

\section{Wavefunctions from conformal blocks}\label{Sec:wfcft}

We first explain how states of spin systems can be constructed from conformal fields. Our starting point is the correlator
\begin{equation}\label{CFTcor}
\langle \phi_{j_1}(z_1,\bar{z}_1) \phi_{j_2}(z_2,\bar{z}_2) \ldots \phi_{j_N}(z_N,\bar{z}_N) \rangle
\end{equation}
of $N$ primary fields $\phi_{j_i}(z_i,\bar{z}_i)$ of a CFT. Here, $\langle\ldots\rangle$ stands for the vacuum expectation value, $z_i$ is a coordinate in the complex plane, and $\bar{z}_i$ is the complex conjugate of $z_i$. The field $\phi_{j_i}(z_i,\bar{z}_i)$ has spin $j_i$, and it therefore has $2j_i+1$ components $\phi_{j_i,s_i}(z_i,\bar{z}_i)$, where $s_i$ labels the $2j_i+1$ possible values of the third component of the spin. It follows that \eref{CFTcor} is a vector with $\prod_{i=1}^N(2j_i+1)$ components.

The correlator \eref{CFTcor} can be broken up into a sum over terms corresponding to different ways of fusing the fields present in the correlator to the vacuum state. Utilizing also that the field $\phi_{j_i}(z_i,\bar{z}_i)$ separates into a holomorphic and an antiholomorphic part, i.e., $\phi_{j_i}(z_i,\bar{z}_i)=\phi_{j_i}(z_i)\otimes\bar{\phi}_{j_i}(\bar{z}_i)$, we can write the components of \eref{CFTcor} as
\begin{equation}\label{corcb}
\langle \phi_{j_1,s_1}(z_1,\bar{z}_1) \ldots \phi_{j_N,s_N}(z_N,\bar{z}_N) \rangle
=\sum_k f_k(s_j,z_j) \overline{f_k(s_j,z_j)},
\end{equation}
where
\begin{equation}\label{cb}
f_k(s_j,z_j)\equiv\langle \phi_{j_1,s_1}(z_1)\phi_{j_2,s_2}(z_2) \cdots \phi_{j_N,s_N}(z_N) \rangle_k
\end{equation}
is a conformal block (also known as a chiral correlator), $k$ labels the different ways of fusing the fields to the vacuum state, and the bar means complex conjugation.

The next step is to regard $f_k(s_j,z_j)$ as the components of a wavefunction $|\psi_k\rangle$ describing the state of $N$ spins with spin quantum numbers $j_1,j_2,\ldots,j_N$. Explicitly,
\begin{equation}\label{psik}
|\psi_k\rangle=
\mathcal{C}_k\sum_{s_1,\ldots,s_N}\psi_k(s_1,\ldots,s_N)|s_1,\ldots,s_N\rangle,
\end{equation}
where
\begin{equation}\label{ccb}
\psi_k(s_1,\ldots,s_N)\propto f_k(s_j,z_j)
\end{equation}
and $\mathcal{C}_k$ is a normalization constant that allows us to choose the normalization of $\psi_k(s_1,\ldots,s_N)$ after convenience. The complex coordinates $z_j=x_j+\rmi y_j$ are fixed parameters of the wavefunction and are naturally interpreted as the physical positions $(x_j,y_j)$ of the spins in the two-dimensional plane. In other words, the choice of $z_j$ defines the lattice under consideration. Note that arbitrary lattices can be considered, since the only restriction on $z_j$ is that $z_j\neq z_l$ for $j\neq l$.

In the present paper, we shall take the CFT to be the WZW model based on the Kac-Moody algebra $SU(2)_1$. This CFT has central charge $c=1$ and two primary fields $\phi_{0}(z)$ and $\phi_{1/2}(z)$ with spin 0 and spin $1/2$, respectively. $\phi_{0}(z)$ has conformal weight $h_0=0$, and $\phi_{1/2}(z)$ has conformal weight $h_{1/2}=1/4$. Since the spin-$0$ primary field is the identity, we shall take all the fields in the correlator to be spin-$1/2$ primary fields. The number of conformal blocks is dictated by the fusion rules
\begin{equation}\label{fusion}
\phi_0\times\phi_0=\phi_0, \quad \phi_0\times\phi_{1/2}=\phi_{1/2}, \quad \phi_{1/2}\times\phi_{1/2}=\phi_0.
\end{equation}
The fields can only fuse to the identity if $N$ is even, and we shall therefore assume this to be the case throughout. The number of conformal blocks on a Riemann surface with genus $g$ is then $2^g$. There is hence one state on the Riemann sphere ($g=0$) and two states on the torus ($g=1$).

The state on the Riemann sphere (complex plane) can be written as \cite{CS,nsc2}
\begin{equation}\label{WFS}
\psi_k(s_1,\ldots,s_N)=\delta_\mathbf{s} \chi_\mathbf{s}
\prod_{i<j}(z_i-z_j)^{(s_is_j+1)/2}\quad(\rm{plane}),
\end{equation}
where $k=0$ can take only one value and $s_i\in\{-1,+1\}$ is defined such that the $z$-component of the spin of the $i$th field is $s_i/2$. The factor
\begin{equation}
\delta_\mathbf{s}=\left\{
\begin{array}{ll}
1 & \rm{for\ }\sum_{i=1}^Ns_i=0\\
0 & \rm{otherwise}
\end{array}\right.
\end{equation}
is the delta function factor, and
\begin{equation}\label{Marshall}
\chi_\mathbf{s}=\prod_{j=1}^N(-1)^{(j-1)(s_j+1)/2}
\end{equation}
is the Marshall sign factor. This formula is the one used to consider lattices with open boundary conditions.

\begin{figure}
\begin{indented}\item[]
\includegraphics[width=0.45\columnwidth]{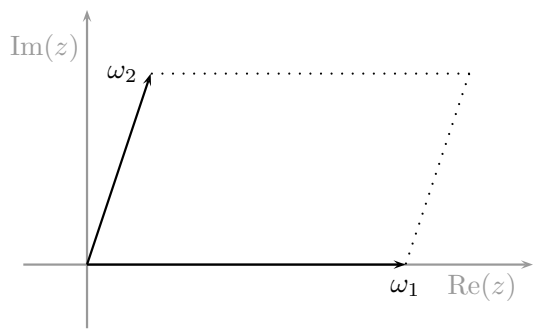}
\caption{Definition of the torus through the periods $\omega_1$ and $\omega_2$. The points $z$ and $z+n\omega_1+m\omega_2$, where $n$ and $m$ are integers, are identified. The coordinate system is chosen such that $\omega_1$ is real and positive and ${\rm Im}(\omega_2)>0$. The modular parameter $\tau$ is defined as $\tau=\omega_2/\omega_1$.}\label{fig:periodic}
\end{indented}
\end{figure}

We shall here mostly be interested in periodic boundary conditions, which corresponds to defining the theory on the torus. The torus is defined by specifying two complex numbers $\omega_1$ and $\omega_2$, which are called the periods of the torus, and then identifying all points in the complex plane that differ by integer multiples of these periods. The $z_i$ in \eref{cb} are then restricted to lie within one parallelogram with sides $\omega_1$ and $\omega_2$ as illustrated in figure \ref{fig:periodic}. $\omega_1$ and $\omega_2$ must be nonzero and have different phases, and we shall choose the coordinate system such that $\omega_1$ is real and positive and $\omega_2$ has positive imaginary part. It is convenient to define the modular parameter $\tau=\omega_2/\omega_1$ and use the scaled coordinates $\zeta_i=z_i/\omega_1$. In terms of these, the states \eref{ccb} take the form
\begin{equation}\label{WFT}
\fl\psi_k(s_1,\ldots,s_N)=\delta_\mathbf{s}\chi_\mathbf{s}
\underbrace{\thetaf{k}{0}{\sum_{i=1}^N \zeta_is_i}{2\tau}}_{\rm Centre\ of\ mass\ factor}
\underbrace{\prod_{i<j}E\left(\zeta_i-\zeta_j,\tau\right)^{(s_is_j+1)/2}}_{\rm Jastrow\ factor}
\qquad(\rm{torus}),
\end{equation}
as we shall derive in the next section. Here, $k\in\{0,1/2\}$ and $s_i\in\{-1,+1\}$. The definition and some properties of the Riemann theta function $\thetaf{a}{b}{\zeta}{\tau}$ can be found in \ref{Sec:AppA}, and
\begin{equation}\label{Edef}
E(\zeta_i-\zeta_j,\tau)\equiv\frac{\theta_1(\zeta_i-\zeta_j,\tau)}
{\partial_\zeta\theta_1(\zeta,\tau)|_{\zeta=0}}, \qquad
\theta_1(\zeta,\tau)\equiv\thetaf{1/2}{1/2}{\zeta}{\tau},
\end{equation}
is the prime form. In the following, we shall refer to the four factors on the right hand side of \eref{WFT} as the delta function factor, the Marshall sign factor, the centre of mass factor, and the Jastrow factor, respectively.

\section{Conformal blocks of the $SU(2)_1$ WZW model on the torus}\label{Sec:cblock}

In this section, we derive \eref{WFT}. We also demonstrate that if the numbering of the spins is altered, then \eref{WFS} and \eref{WFT} stay the same except that they are multiplied by the sign of the permutation needed to go from one numbering to the other. The choice of numbering is thus not important.

\subsection{CFT correlators}

We first need an expression for the correlator \eref{CFTcor}. To get this, we utilize the fact that the $SU(2)_1$ WZW model can be bosonized in terms of a massless free scalar field $\varphi(\zeta,\bar{\zeta})=\varphi(\zeta)+\bar{\varphi}(\bar{\zeta})$ compactified on a circle of radius $R=\sqrt{2}$ \cite{yellow}. Specifically, $\phi_{j_i,s_i}(\zeta_i,\bar{\zeta}_i)=\phi_{j_i,s_i}(\zeta_i) \otimes\bar{\phi}_{j_i,s_i}(\bar{\zeta}_i)$ can be expressed as the vertex operator
\begin{equation}\label{bosonization}
\phi_{j_i,s_i}(\zeta_i,\bar{\zeta}_i)=
\; :\rme^{\rmi s_i\varphi(\zeta_i,\bar{\zeta}_i)/\sqrt{2}}:\;
=\; :\rme^{\rmi s_i\varphi(\zeta_i)/\sqrt{2}+\rmi s_i\bar{\varphi}(\bar{\zeta}_i)/\sqrt{2}}:\,,
\end{equation}
where $:\ldots:$ denotes normal ordering. We use here the scaled coordinates $\zeta_i$, but we note that this changes the correlator \eref{CFTcor} only by a constant factor. Note that the conformal weight of $:\rme^{\rmi s_i\varphi(\zeta_i)/\sqrt{2}}:$ is $s_i^2/4=1/4$, as it should be.

We can now use the expression
\begin{equation}\label{fullcor}
\left\langle \prod_{i=1}^N :\rme^{\rmi\nu_i\varphi(\zeta_i)+\rmi\bar{\nu}_i \bar{\varphi} (\bar{\zeta_i})}:\right\rangle=
\frac{\delta_\mathbf{\nu}\delta_{\overline{\mathbf{\nu}}}}{|\eta(\tau)|^2}\sum_{(p,\bar{p})\in\Gamma}
A_0^p(\zeta_i,\nu_i) \overline{A_0^p(\zeta_i,\nu_i)}
\end{equation}
for the correlator of a product of generic vertex operators on the torus derived in \cite{DVV} (see also \cite{alvarez}). Here, $(\nu_i,\bar{\nu}_i)\in\Gamma$, $\Gamma$ is the lattice of momenta given by
\begin{equation}
p=\frac{n}{R}+\frac{1}{2}mR, \quad
\bar{p}=\frac{n}{R}-\frac{1}{2}mR, \qquad n,m \in \mathbb{Z},
\end{equation}
$\delta_\mathbf{\nu}\delta_{\overline{\mathbf{\nu}}}$ is one for $\sum_i\nu_i=\sum_i\bar{\nu}_i=0$ and zero otherwise,
\begin{equation}\label{dedekind}
\eta(\tau)=\rme^{\pi\rmi\tau/12}\prod_{n=1}^\infty(1-\rme^{2\pi\rmi\tau n})
\end{equation}
is the Dedekind eta function,
\begin{equation}\label{A}
A_0^p(\zeta_i,\nu_i)=\rme^{\rmi\pi p^2\tau+2\pi\rmi p\sum_i \zeta_i \nu_i}
\prod_{i<j}E(\zeta_i-\zeta_j,\tau)^{\nu_i\nu_j},
\end{equation}
and the bar means complex conjugation and transformation of $\nu_i$ and $p$ into $\bar{\nu}_i$ and $\bar{p}$.

For $R=\sqrt{2}$, the momenta become
\begin{equation}
p=\frac{1}{\sqrt{2}}(n+m),\quad\bar{p}=\frac{1}{\sqrt{2}}(n-m),\qquad
n,m\in \mathbb{Z}.
\end{equation}
Noting that $n+m$ always has the same parity as $n-m$, we can write $n+m=2r+2k$ and $n-m=2s+2k$ and replace the sums over $n$ and $m$ by a sum over $k\in\{0,1/2\}$ and sums over $r\in\mathbb{Z}$ and $s\in\mathbb{Z}$. Therefore
\begin{equation}\label{resum}
\fl\sum_{(p,\bar{p})\in\Gamma}
A_0^p(\zeta_i,\nu_i) \overline{A_0^p(\zeta_i,\nu_i)}
=\sum_{k\in\{0,1/2\}}\sum_{r\in\mathbb{Z}}
A_0^{\sqrt{2}(r+k)}(\zeta_i,\nu_i) \overline{\sum_{s\in\mathbb{Z}}A_0^{\sqrt{2}(s+k)}(\zeta_i,\nu_i)}.
\end{equation}
Combining \eref{corcb}, \eref{bosonization}, \eref{fullcor}, \eref{A}, \eref{resum}, and \eref{A1}, we conclude that the conformal blocks are
\begin{equation}\label{cbexp}
f_k(\zeta_i,s_i)=\frac{\xi_k\delta_\mathbf{s}}{\eta(\tau)}
\thetaf{k}{0}{\sum_i \zeta_is_i}{2\tau}
\prod_{i<j}E\left(\zeta_i-\zeta_j,\tau\right)^{s_is_j/2} ,
\end{equation}
where $k\in\{0,1/2\}$ and $\xi_k$ is a phase factor that cannot be determined from the above arguments. $\xi_k$ can be written as $\xi_k=\rme^{\rmi g_k}$, where $g_k$ is a real-valued function. Since $g_k$ must also be holomorphic in $\zeta_i$ to ensure that $f_k$ is holomorphic, we conclude that $\xi_k$ cannot depend on $\zeta_i$. It may, however, depend on $s_i$.

\subsubsection{Singlet property}\label{Sec:singlet}

The dependence of the phase factors $\xi_k$ on $s_i$ can be determined from the $SU(2)$ symmetry of the CFT, which dictates that the wavefunction must be a spin-$0$ state. Since $\xi_k$ does not depend on $\zeta_i$, it is sufficient to determine its value for a convenient choice of $\zeta_i$. The idea in the following is to utilize that the torus locally looks like the plane. We can therefore find the phases by putting all the spins close together and comparing the conformal blocks to \eref{WFS}. Let us specifically put all $\zeta_i$ close to the origin, such that $|\zeta_i|\leq\epsilon$ for all $i$, where $\epsilon$ is a small positive number. We shall choose $\epsilon$ to be small enough that $N^2\epsilon^2\ll1$, $\left|\sum_i \zeta_is_i\right|\ll\mathrm{Im}(\tau)^{1/2}$, and $|\zeta_i-\zeta_j|\ll\mathrm{Im}(\tau)^{1/2}$ for all $i$ and $j$. Then
\begin{equation}
\fl\thetaf{k}{0}{\sum_i \zeta_is_i}{2\tau}=
\sum_{n \in \mathbb{Z}} \rme^{2\pi \rmi\tau(n+k)^2
+2\pi \rmi(n+k)\sum_i \zeta_is_i}
=\sum_{n \in \mathbb{Z}} \rme^{2\pi \rmi\tau(n+k)^2}
+\mathcal{O}\left(\epsilon^2\right)
\end{equation}
and
\begin{eqnarray}
\fl E(\zeta_i-\zeta_j,\tau)=\frac{1}{\partial_ \zeta\theta_1(\zeta,\tau)|_{\zeta=0}}
\sum_{n \in \mathbb{Z}} \rme^{\pi \rmi\tau\left(n+\frac{1}{2}\right)^2
+2\pi \rmi\left(n+\frac{1}{2}\right)\left(\zeta_i-\zeta_j +\frac{1}{2}\right)}\nonumber\\
=\frac{1}{\partial_ \zeta\theta_1(\zeta,\tau)|_{\zeta=0}}\sum_{n \in \mathbb{Z}} 2\pi \rmi\left(n+\frac{1}{2}\right)(\zeta_i-\zeta_j)
\rme^{\pi \rmi\tau\left(n+\frac{1}{2}\right)^2+\pi \rmi\left(n+\frac{1}{2}\right)}
+\mathcal{O}\left(\epsilon^3\right)\nonumber\\
=\zeta_i-\zeta_j+\mathcal{O}\left(\epsilon^3\right).
\end{eqnarray}
Inserting these expressions in \eref{cbexp}, we get
\begin{equation}\label{auxf}
f_k(\zeta_i,s_i)\propto\xi_k\delta_{\mathbf{s}}\prod_{i<j}(\zeta_i-\zeta_j)^{s_is_j/2}
(1+\mathcal{O}(\epsilon^2)).
\end{equation}
Since $\sum_{i<j}s_is_j=\frac{1}{2}\sum_{i\neq j}s_is_j =-\frac{1}{2}\sum_{i=1}^Ns_is_i=-\frac{N}{2}$ is a constant for $\sum_{j=1}^Ns_j=0$, we can replace $(\zeta_i-\zeta_j)$ by $(z_i-z_j)$ in \eref{auxf}. Comparing \eref{auxf} to the $SU(2)$ invariant state on the complex plane \eref{WFS}, we observe that $\xi_k$ is the Marshall sign factor defined in \eref{Marshall}, provided we choose the phase convention of the square roots present in the Jastrow factor such that $(\zeta_i-\zeta_j)^{-1/2}=((\zeta_i-\zeta_j)^{1/2})^{-1}$. Up to a constant phase factor, we thus have
\begin{equation}\label{cbfinal}
f_k(\zeta_i,s_i)=\frac{\delta_\mathbf{s}\chi_\mathbf{s}}{\eta(\tau)}
\thetaf{k}{0}{\sum_i \zeta_is_i}{2\tau}
\prod_{i<j}E\left(\zeta_i-\zeta_j,\tau\right)^{s_is_j/2},
\end{equation}
where $E(\zeta_i-\zeta_j,\tau)^{-1/2}=(E(\zeta_i-\zeta_j,\tau)^{1/2})^{-1}$ is assumed. As far as the wavefunction is concerned, we are free to multiply the conformal block by an $s_i$-independent factor. To get \eref{WFT}, we multiply \eref{cbfinal} by $\eta(\tau)\prod_{i<j}E(\zeta_i-\zeta_j,\tau)^{1/2}$. Note that the latter factor changes the exponent in the Jastrow factor from the half-integer $s_is_j/2$ to the integer $(s_is_j+1)/2$ so that the need to take square roots is eliminated. As a final remark, we note that there is a connection between the singlet property of the wavefunctions and Fay's trisecant identity as is clear from the alternative derivation of the singlet property for four spins given in \ref{Sec:AppB}.

\subsection{Independence of the ordering of the spins}\label{Sec:numb}

Let us also demonstrate that the state \eref{WFT} is independent (up to an overall sign factor) of the chosen ordering of the spins. Let $p$ be a bijective map from $\{1,2,\ldots,N\}$ to $\{1,2,\ldots,N\}$, which defines an alternative ordering. Using this ordering, the state \eref{WFT} takes the form
\begin{eqnarray}
\fl\psi_k^{(p)}(s_1,\ldots,s_N)=&
\delta_\mathbf{s}
\prod_{j=1}^N(-1)^{(p(j)-1)(s_j+1)/2}
\thetaf{k}{0}{\sum_{i=1}^N \zeta_is_i}{2\tau}\nonumber\\
&\times\prod_{\{i,j\,\in\{1,2,\ldots,N\}|p(i)<p(j)\}}
E(\zeta_i-\zeta_j,\tau)^{(s_is_j+1)/2}.\label{psip}
\end{eqnarray}
Let us specifically consider the transposition
\begin{equation}\label{number}
p(j)=\left\{
\begin{array}{ll}
j & {\rm for\ } j\neq j_1,j_2\\
j_2 & {\rm for\ } j=j_1\\
j_1 & {\rm for\ } j=j_2
\end{array}
\right.,
\end{equation}
where $j_1$ and $j_2$ are two numbers in $\{1,2,\ldots,N\}$ satisfying $j_1<j_2$. Since $E$ is an odd function in the first argument according to \eref{A11}, we get
\begin{eqnarray}\label{swap}
\fl\frac{\psi_k^{(p)}(s_1,\ldots,s_N)}{\psi_k(s_1,\ldots,s_N)}
=\frac{(-1)^{(j_2-1)(s_{j_1}+1)/2+(j_1-1)(s_{j_2}+1)/2}}
{(-1)^{(j_2-1)(s_{j_2}+1)/2+(j_1-1)(s_{j_1}+1)/2}}
(-1)^{(s_{j_1}s_{j_2}+1)/2+\sum_{j=j_1+1}^{j_2-1}(s_{j_1}s_j+s_{j_2}s_j+2)/2}\nonumber\\
=(-1)^{(j_2-j_1)(s_{j_1}-s_{j_2})/2}
(-1)^{(s_{j_1}s_{j_2}+1)/2+\sum_{j=j_1+1}^{j_2-1}s_j(s_{j_1}+s_{j_2})/2
+j_2-j_1-1}\nonumber\\
=(-1)^{(j_2-j_1)(s_{j_1}-s_{j_2})/2+(s_{j_1}s_{j_2}+1)/2+(j_2-j_1-1)(s_{j_1}+s_{j_2})/2
+j_2-j_1-1}\nonumber\\
=(-1)^{(j_2-j_1)(s_{j_1}+1)+(s_{j_1}-1)(s_{j_2}-1)/2-1}=-1.
\end{eqnarray}
Swapping two labels in the ordering sequence thus only changes the overall sign of the wavefunction. Since any permutation $p$ can be decomposed into a product of transpositions \eref{number}, we conclude that the choice of ordering is unimportant. The exact same arguments apply to \eref{WFS}, because $z_i-z_j$ is an odd function of $z_i-z_j$.

\section{Connection to the KL states on the torus}\label{Sec:KL}

The CFT states can be transformed into states describing particles on a lattice by regarding all spins with $s_i=+1$ as occupied sites and all spins with $s_i=-1$ as empty sites. Doing so, we demonstrate in this section that the CFT states on the torus \eref{WFT} coincide with the KL states on the torus if the spins sit on an $L_x\times L_y$ square lattice with lattice constant $\sqrt{2\pi}$ and $\sum_j\zeta_j=0$. The CFT states thus provide a natural generalization of the KL states to arbitrary lattices.

\subsection{Statement of the result}

Expressed mathematically, the desired square lattice is obtained by choosing \begin{equation}
\omega_1=\sqrt{2\pi}L_x,\quad \omega_2=\rmi\sqrt{2\pi}L_y,\quad \tau=\rmi L_y/L_x, \end{equation}
and
\begin{equation}\label{coor}
z_{n+mL_x+1}=\sqrt{2\pi}[n-l_x+\rmi(m-l_y)],
\end{equation}
where
\begin{eqnarray}
l_x=(L_x-1)/2,\quad &n\in\{0,1,\ldots,L_x-1\},\\
l_y=(L_y-1)/2, \quad &m\in\{0,1,\ldots,L_y-1\}.
\end{eqnarray}
The result of this section is that
\begin{equation}\label{wfc}
\psi_k(s_1,\ldots,s_N)=(-1)^{2k}c\psi_k(Z_1,\ldots,Z_{N/2})\prod_iG_k(Z_i),
\quad k=0,1/2.
\end{equation}
The right hand side of this expression is the KL state on the torus. Specifically, $\psi_k(Z_1,\ldots,Z_{N/2})$ is the Laughlin state with Landau level filling factor $1/2$ of $N/2$ particles on the torus \cite{L89,torus1,torus2,RZ96}
\begin{eqnarray}
\fl\psi_k(Z_1,\ldots,Z_{N/2})=\nonumber\\
\thetaf{N/2-k}{-(N-1)}{\frac{2\sum_iZ_i}{\sqrt{2\pi}L_x}}{2\tau} \prod_{i<j}\left[\thetaf{1/2}{1/2}{\frac{Z_i-Z_j}{\sqrt{2\pi}L_x}}{\tau}\right]^2 \rme^{-\frac{1}{2}\sum_iY_i^2}\label{wfp}
\end{eqnarray}
with the particle positions $Z_i=X_i+\rmi Y_i$ restricted to the considered square lattice, i.e., $Z_i\in\{z_1,\ldots,z_N\}$. Particles are identified with spins in the `up' state such that $\{Z_1,\ldots,Z_{N/2}\}=\{z_i|i\in\{1,2,\ldots,N\} \wedge s_i=1\}$. As in \eref{KLsphere}, $\prod_iG_k(Z_i)$ is a gauge factor, which can only take values in the set $\{-1,+1\}$, and the purpose of this factor is to ensure that the right hand side of \eref{wfc} is a singlet. Finally, $c$ is an overall constant that only depends on the choice of $L_x$ and $L_y$ and is unimportant, since it is absorbed in the normalization of the state anyway. We find specifically that
\begin{equation}
\prod_{i=1}^{N/2}G_k(Z_i)= \left\{\begin{array}{ll}
\prod_{i=1}^N
(-1)^{m_iq_i} & {\rm for\ }L_y{\rm\ even}\\
\prod_{i=1}^N
(-1)^{(n_i+m_i)q_i} & {\rm for\ }L_y{\rm\ odd}\\ \end{array}\right..
\end{equation}
Here, $n_i={\rm Re}(z_i)/\sqrt{2\pi}+l_x$ and $m_i={\rm Im}(z_i)/\sqrt{2\pi}+l_y$ as in \eref{coor} and
\begin{equation}\label{qi}
q_i=(s_i+1)/2
\end{equation}
such that only sites with $s_i=+1$ contribute to the products. Note that $q_i$ is the occupation number on site $i$ since $q_i=1$ for spin up and $q_i=0$ for spin down. In the following, we derive \eref{wfc} by rewriting each of the factors in \eref{WFT}.

\begin{figure}
\begin{indented}\item[]
\includegraphics[width=0.7\columnwidth]{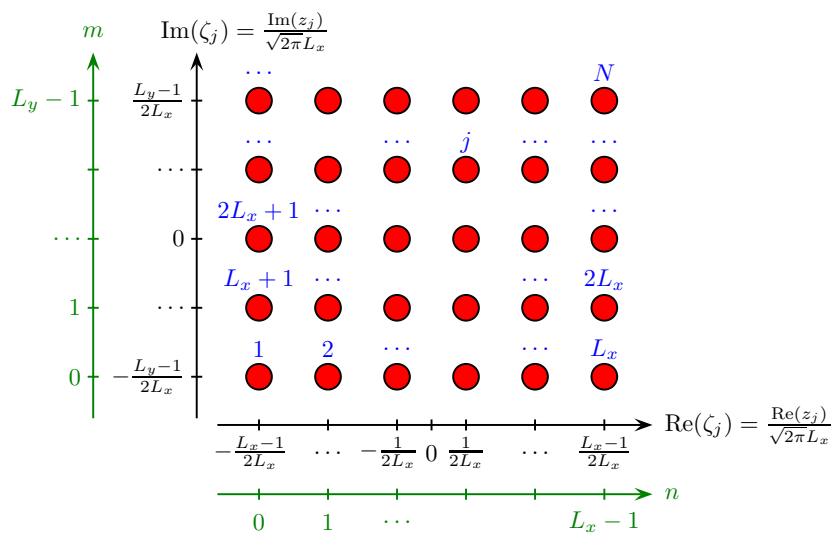}
\caption{We use four different ways to label the lattice sites. $(n,m)$ is the coordinates on the lattice with $n=0,1,\ldots,L_x-1$ and $m=0,1,\ldots,L_y-1$. $j=n+mL_x+1$ is the site index, which numbers the sites from $1$ to $N$ as shown with the labels above each site. $z_j$ defined in \eref{coor} is the position in the complex plane of the site with index $j$ (in units of the magnetic length, when comparing to FQH states), and $\zeta_j=z_j/\omega_1$, where $\omega_1=\sqrt{2\pi}L_x$ is the width of the lattice in the $x$-direction. \label{Fig:sqlat}}
\end{indented}
\end{figure}

\subsection{The delta function factor}

The condition $\sum_{i=1}^Ns_i=0$ translates into $\sum_{i=1}^Nq_i=N/2$, and the delta function factor thus ensures that the lattice is half filled when a spin up is considered as an occupied site and a spin down is considered as an empty site. In other words, the delta function factor ensures that there are $N/2$ of the $Z_i$ coordinates.

\subsection{The Marshall sign factor}

The Marshall sign factor can be rewritten into
\begin{equation}\label{msfq}
\chi_\mathbf{s}=\prod_{j=1}^N(-1)^{(j-1)(s_j+1)/2}=\prod_{j=1}^N(-1)^{(j-1)q_j}.
\end{equation}

\subsection{The centre of mass factor}

Let us first note that
\begin{equation}
\sum_{i=1}^Ns_iz_i=2\sum_{i=1}^Nq_iz_i-\sum_{i=1}^Nz_i
=2\sum_{i=1}^Nq_iz_i=2\sum_{i=1}^{N/2}Z_i.
\end{equation}
Using \eref{A2}, \eref{A3}, and the assumption that $N$ is even, it then follows that
\begin{equation}\label{cmfq}
\thetaf{k}{0}{\frac{\sum_i z_is_i}{\sqrt{2\pi}L_x}}{2\tau}=(-1)^{2k} \thetaf{N/2-k}{-(N-1)}{\frac{2\sum_iZ_i}{\sqrt{2\pi}L_x}}{2\tau}.
\end{equation}

\subsection{The Jastrow factor}

We write
\begin{equation}\label{jfq1}
\fl\prod_{i<j}E(\zeta_i-\zeta_j,\tau)^{(s_is_j+1)/2}=
\prod_{i<j}E(\zeta_i-\zeta_j,\tau)
\prod_{i<j}E(\zeta_i-\zeta_j,\tau)^{2q_iq_j}
\prod_{i<j}E(\zeta_i-\zeta_j,\tau)^{-q_i-q_j}.
\end{equation}
The first factor on the right hand side is an overall unimportant factor, the second factor appears in \eref{wfp}, and the third factor is to be compared to the Gaussian factor in \eref{wfp}. Since $E$ is an odd function in the first argument according to \eref{A11}, we have
\begin{eqnarray}\label{jfq2}
\fl\prod_{i<j}E(\zeta_i-\zeta_j,\tau)^{-q_i-q_j}
=\prod_{i<j}E(\zeta_i-\zeta_j,\tau)^{-q_i}
\prod_{j<i}E(\zeta_j-\zeta_i,\tau)^{-q_i}\nonumber\\
=\prod_{j<i}(-1)^{-q_i}\prod_{i\neq j} E(\zeta_i-\zeta_j,\tau)^{-q_i}
=\prod_{j=1}^N(-1)^{-(j-1)q_j}\prod_{i=1}^N f(\zeta_i)^{q_i},
\end{eqnarray}
where
\begin{eqnarray}
f(\zeta_i)\equiv\prod_{j(\neq i)} E(\zeta_i-\zeta_j,\tau)^{-1}
\end{eqnarray}
and the notation $\prod_{j(\neq i)}$ means the product over all $j$ except $i$ and no product over $i$.

We next evaluate $f(\zeta_i)$. For $m\neq L_y-1$, we find
\begin{eqnarray}
\fl\frac{f(\zeta_{n+(m+1)L_x+1})}{f(\zeta_{n+mL_x+1})}=
\frac{\prod_{{0\leq n'\leq L_x-1\atop 0\leq m'\leq L_y-1}\atop n'+\rmi m'\neq n+\rmi m} E\left(\frac{n+\rmi m-n'-\rmi m'}{L_x},\tau\right)} {\prod_{{0\leq n'\leq L_x-1\atop 0\leq m'\leq L_y-1}\atop n'+\rmi m'\neq n+\rmi(m+1)} E\left(\frac{n+\rmi(m+1)-n'-\rmi m'}{L_x},\tau\right)}\nonumber\\
=\frac{\prod_{{0\leq n'\leq L_x-1\atop 0\leq m'\leq L_y-1}\atop n'+\rmi m'\neq n+\rmi m}E\left(\frac{n+\rmi m-n'-\rmi m'}{L_x},\tau\right)} {\prod_{{0\leq n'\leq L_x-1\atop -1\leq m'\leq L_y-2}\atop n'+\rmi m'\neq n+\rmi m} E\left(\frac{n+\rmi m-n'-\rmi m'}{L_x},\tau\right)}
=\frac{\prod_{0\leq n'\leq L_x-1} E\left(\frac{n+\rmi m-n'-\rmi(L_y-1)}{L_x},\tau\right)} {\prod_{0\leq n'\leq L_x-1} E\left(\frac{n+\rmi m-n'+\rmi}{L_x},\tau\right)}\nonumber\\
=\prod_{0\leq n'\leq L_x-1} \rme^{\rmi\pi-\rmi\pi\tau}\rme^{2\pi \rmi[n+\rmi m-n'+\rmi]/L_x}
=\rme^{\rmi\pi L_x-\rmi\pi\tau L_x}\rme^{2\pi \rmi[n+\rmi m-(L_x-1)/2+\rmi]}\nonumber\\
=-\rme^{\pi L_y}\rme^{-2\pi (m+1)}
=-\rme^{-2\pi (m-l_y)-\pi},
\end{eqnarray}
where we have used \eref{A13}. For $m=L_y-1$, the relevant quantity is
\begin{eqnarray}
\fl\frac{f(\zeta_{n+1})}{f(\zeta_{n+(L_y-1)L_x+1})}=
\frac{\prod_{{0\leq n'\leq L_x-1\atop 0\leq m'\leq L_y-1}\atop n'+\rmi m'\neq n+\rmi(L_y-1)} E\left(\frac{n+\rmi(L_y-1)-n'-\rmi m'}{L_x},\tau\right)} {\prod_{{0\leq n'\leq L_x-1\atop 0\leq m'\leq L_y-1}\atop n'+\rmi m'\neq n} E\left(\frac{n-n'-\rmi m'}{L_x},\tau\right)}\nonumber\\
=\frac{\prod_{{0\leq n'\leq L_x-1\atop 0\leq m'\leq L_y-1}\atop n'+\rmi m'\neq n+\rmi(L_y-1)} E\left(\frac{n+\rmi(L_y-1)-n'-\rmi m'}{L_x},\tau\right)} {\prod_{{0\leq n'\leq L_x-1\atop 0\leq m'\leq L_y-1}\atop n'+\rmi m'\neq n} \left[E\left(\frac{n+\rmi L_y-n'-\rmi m'}{L_x},\tau\right)
\rme^{\rmi\pi-\rmi\pi\tau+2\pi \rmi(n+\rmi L_y-n'-\rmi m')/L_x}\right]}\nonumber\\
=\frac{\prod_{{0\leq n'\leq L_x-1\atop 0\leq m'\leq L_y-1}\atop n'+\rmi m'\neq n+\rmi(L_y-1)} E\left(\frac{n+\rmi(L_y-1)-n'-\rmi m'}{L_x},\tau\right)} {\prod_{{0\leq n'\leq L_x-1\atop -1\leq m'\leq L_y-2}\atop n'+\rmi m'\neq n-\rmi} \left[E\left(\frac{n+\rmi(L_y-1)-n'-\rmi m'}{L_x},\tau\right)
\rme^{\rmi\pi-\rmi\pi\tau+2\pi \rmi(n+\rmi(L_y-1)-n'-\rmi m')/L_x}\right]}\nonumber\\
=\frac{\prod_{0\leq n'\leq L_x-1\atop n'\neq n} E\left(\frac{n-n'}{L_x},\tau\right)} {\prod_{0\leq n'\leq L_x-1\atop n'\neq n} E\left(\frac{n-n'}{L_x}+\tau,\tau\right)}
\times\frac{\rme^{-\rmi\pi(N-1)+\rmi\pi\tau(N+1)}} {\prod_{0\leq n'\leq L_x-1\atop -1\leq m'\leq L_y-2} \rme^{2\pi \rmi(n+\rmi(L_y-1)-n'-\rmi m')/L_x}}\nonumber\\
=-\frac{\prod_{0\leq n'\leq L_x-1\atop n'\neq n} E\left(\frac{n-n'}{L_x},\tau\right)}
{\prod_{0\leq n'\leq L_x-1\atop n'\neq n} \left[E\left(\frac{n-n'}{L_x},\tau\right)\rme^{-\rmi\pi-\rmi\pi\tau-2\pi \rmi(n-n')/L_x}\right]}
\times\frac{\rme^{\rmi\pi\tau(N+1)}} {\rme^{2\pi \rmi(n-l_x+\rmi(L_y+1)/2)L_y}}\nonumber\\
=\prod_{0\leq n'\leq L_x-1} \rme^{\rmi\pi+\rmi\pi\tau+2\pi \rmi(n-n')/L_x}
\times\frac{\rme^{\rmi\pi\tau N}} {\rme^{2\pi \rmi(n-l_x+\rmi(L_y+1)/2)L_y}}
=-(-1)^{L_y}.
\end{eqnarray}
For $n\neq L_x-1$, we get
\begin{eqnarray}
\fl\frac{f(\zeta_{n+1+mL_x+1})}{f(\zeta_{n+mL_x+1})}=
\frac{\prod_{{0\leq n'\leq L_x-1\atop 0\leq m'\leq L_y-1}\atop n'+\rmi m'\neq n+\rmi m} E\left(\frac{n+\rmi m-n'-\rmi m'}{L_x},\tau\right)} {\prod_{{0\leq n'\leq L_x-1\atop 0\leq m'\leq L_y-1}\atop n'+\rmi m'\neq n+1+\rmi m} E\left(\frac{n+1+\rmi m-n'-\rmi m'}{L_x},\tau\right)}
=\frac{\prod_{{0\leq n'\leq L_x-1\atop 0\leq m'\leq L_y-1}\atop n'+\rmi m'\neq n+\rmi m} E\left(\frac{n+\rmi m-n'-\rmi m'}{L_x},\tau\right)} {\prod_{{-1\leq n'\leq L_x-2\atop 0\leq m'\leq L_y-1}\atop n'+\rmi m'\neq n+\rmi m} E\left(\frac{n+\rmi m-n'-\rmi m'}{L_x},\tau\right)}\nonumber\\
=\frac{\prod_{0\leq m'\leq L_y-1} E\left(\frac{n+\rmi m-(L_x-1)-\rmi m'}{L_x},\tau\right)} {\prod_{0\leq m'\leq L_y-1} E\left(\frac{n+\rmi m+1-\rmi m'}{L_x},\tau\right)}
=(-1)^{L_y},
\end{eqnarray}
where we used \eref{A12} in the last line. For $n=L_x-1$, the relevant quantity is
\begin{eqnarray}
\fl\frac{f(\zeta_{mL_x+1})}{f(\zeta_{L_x-1+mL_x+1})}
=\frac{\prod_{{0\leq n'\leq L_x-1\atop 0\leq m'\leq L_y-1}\atop n'+\rmi m'\neq L_x-1+\rmi m} E\left(\frac{L_x-1+\rmi m-n'-\rmi m'}{L_x},\tau\right)} {\prod_{{0\leq n'\leq L_x-1\atop 0\leq m'\leq L_y-1}\atop n'+\rmi m'\neq \rmi m} E\left(\frac{\rmi m-n'-\rmi m'}{L_x},\tau\right)}\nonumber\\
=-\frac{\prod_{{0\leq n'\leq L_x-1\atop 0\leq m'\leq L_y-1}\atop n'+\rmi m'\neq L_x-1+\rmi m} E\left(\frac{L_x-1+\rmi m-n'-\rmi m'}{L_x},\tau\right)} {\prod_{{-1\leq n'\leq L_x-2\atop 0\leq m'\leq L_y-1}\atop n'+\rmi m'\neq -1+\rmi m} E\left(\frac{L_x-1+\rmi m-n'-\rmi m'}{L_x},\tau\right)}\nonumber\\
=-\frac{\prod_{0\leq m'\leq L_y-1\atop m'\neq m} E\left(\frac{\rmi m-\rmi m'}{L_x},\tau\right)} {\prod_{0\leq m'\leq L_y-1\atop m'\neq m} E\left(\frac{L_x+\rmi m-\rmi m'}{L_x},\tau\right)}
=(-1)^{L_y}.
\end{eqnarray}
The solution of these equations is
\begin{equation}\label{fz}
f(\zeta_{n+mL_x+1})\propto \left\{\begin{array}{ll}
(-1)^{m}\rme^{-\pi (m-l_y)^2} & {\rm for\ }L_y{\rm\ even}\\
(-1)^{n+m}\rme^{-\pi (m-l_y)^2} & {\rm for\ }L_y{\rm\ odd}
\end{array}\right..
\end{equation}
The result \eref{wfc} is then obtained by combining the delta function factor, \eref{msfq}, \eref{cmfq}, \eref{jfq1}, \eref{jfq2}, and \eref{fz}.

\section{Transformation properties of the CFT states on a square lattice on the torus}\label{Sec:trans}

In this section, we investigate how the CFT states on the torus transform under different operations. This reveals the symmetries of the states and has the additional advantage of enabling us in certain cases to find linear combinations of the two states on the torus that are necessarily orthogonal. We assume throughout that the spins sit on an $L_x\times L_y$ square lattice as defined in \eref{coor}, but unless it is required for the considered transformation to make sense, we shall not put other restrictions on $L_x$ and $L_y$ than demanding $N=L_xL_y$ to be even. For the sake of generality, we shall allow for twisted boundary conditions in the following. If the twist angles are $\theta_x$ and $\theta_y$, this means that the state acquires a phase of $\rme^{\rmi \theta_x N/2}$ if the lattice is displaced by $\sqrt{2\pi}L_x$ in the $x$-direction and acquires a phase of $\rme^{\rmi \theta_y N/2}$ if the lattice is displaced by $\sqrt{2\pi}L_y$ in the $y$-direction. It is already known how to take the twists into account for the Laughlin states on the torus (see e.g.\ (5.5) in \cite{RZ96}), and by comparing the CFT states to this expression, we conclude that
\begin{eqnarray}\label{WFtwist}
\fl\psi_{a,b}(s_1,\ldots,s_N)=\delta_\mathbf{s}
\chi_\mathbf{s}\thetaf{a}{b}{\sum_{i=1}^N \zeta_is_i}{2\tau}
\prod_{i<j}E\left(\zeta_i-\zeta_j,\tau\right)^{(s_is_j+1)/2}
\quad(\rm{twist})
\end{eqnarray}
provides the desired generalization, where
\begin{equation}\label{ab}
a=k+\frac{\theta_x}{4\pi}, \qquad b=-\frac{\theta_y}{2\pi},
\qquad k\in\{0,1/2\}, \qquad \sum_{j=1}^N\zeta_j=0.
\end{equation}
In analogy to \eref{psik}, we also define
\begin{equation}\label{psiab}
|\psi_{a,b}\rangle=\mathcal{C}_{a,b} \sum_{s_1,\ldots,s_N}\psi_{a,b}(s_1,\ldots,s_N)|s_1,\ldots,s_N\rangle,
\end{equation}
where $\mathcal{C}_{a,b}$ is the normalization constant.

The results of this section obtained for $\theta_x=\theta_y=0$ are summarized in table \ref{Tab:eigen}. From this table we conclude that $\psi_0=\psi_{0,0}(s_1,\ldots,s_N)$ and $\psi_{1/2}=\psi_{1/2,0}(s_1,\ldots,s_N)$ are orthogonal for even-by-odd lattices, whereas $\psi_0+\psi_{1/2}$ and $\psi_0-\psi_{1/2}$ are orthogonal for odd-by-even lattices. For $L_x=L_y$, it is the combinations $\psi_{0}-(1-\sqrt{2})\psi_{1/2}$ and $\psi_{0}-(1+\sqrt{2})\psi_{1/2}$ that are orthogonal. We also note that the operators describing translation in the $x$-direction and in the $y$-direction can be diagonalized simultaneously, which is not true in the continuum case \cite{torus1}. This is so because the area of each lattice site is $2\pi$, and the magnetic flux penetrating an area of $2\pi$ in the fractional quantum Hall setting is one flux quantum. The operator $(T_y)^{-1}(T_x)^{-1}T_yT_x$ thus moves all the particles (or spin ups) around one flux quantum, and therefore the Aharonov-Bohm phase picked up is a multiple of $2\pi$. The transformation properties for general $\theta_x$ and $\theta_y$ can be found in \eref{F}, \eref{Tx}, \eref{Ty}, \eref{R90}, \eref{R180}, \eref{MxT}, and \eref{MyT}. Note also \eref{tildeTx} and \eref{tildeTy} for the transformation properties of a transformed set of translation operators.

\begin{table}
\caption{\label{Tab:eigen} Eigenstates and eigenvalues in the subspace spanned by $\psi_0=\psi_{0,0}(s_1,\ldots,s_N)$ and $\psi_{1/2}=\psi_{1/2,0}(s_1,\ldots,s_N)$ of the following operators: spin flip ($F$), translation by one lattice constant in the $x$-direction ($T_x$), translation by one lattice constant in the $y$-direction ($T_y$), rotation by $90^{\circ}$ ($R_{90^\circ}$), rotation by $180^{\circ}$ ($R_{180^\circ}$), time reversal followed by reflection in the $x$-direction ($M_x\Theta$), and time reversal followed by reflection in the $y$-direction ($M_y\Theta$). We consider four different types of square lattices on the torus: `equal' stands for $L_x=L_y$ with $L_x$ even, `e$\times$e' means $L_x$ even and $L_y$ even, `e$\times$o' means $L_x$ even and $L_y$ odd, and `o$\times$e' means $L_x$ odd and $L_y$ even (odd-by-odd is not allowed since $N=L_xL_y$ must be even). Note that $\psi_0$ and $\psi_{1/2}$ are defined as in \eref{WFT} and are hence not normalized.}
\lineup
\begin{tabular}{@{}lllllllll}
\br
Lattice & Eigenstates & $F$ & $T_x$ & $T_y$ & $R_{90^\circ}$ & $R_{180^\circ}$ & $M_x\Theta$ & $M_y\Theta$\\
\mr
equal & $\psi_{0}-(1-\sqrt{2})\psi_{1/2}$ & $1$ & $1$ & $1$ & $+1$ & $1$ & $1$ & $1$\\
& $\psi_{0}-(1+\sqrt{2})\psi_{1/2}$ & $1$ & $1$ & $1$ & $-1$ & $1$ & $1$ & $1$\\
\ms
e$\times$e & $\psi_0$ & $1$ & $1$ & $1$ & \m- & $1$ & $1$ & $1$\\
& $\psi_{1/2}$ & $1$ & $1$ & $1$ & \m- & $1$ & $1$ & $1$\\
\ms
e$\times$o & $\psi_0$ & $(-1)^{\frac{N}{2}}$ & $(-1)^{\frac{L_x}{2}}$ & $1$ & \m - & $(-1)^{\frac{N}{2}}$ & $1$ & $1$\\
& $\psi_{1/2}$ & $(-1)^{\frac{N}{2}}$ & $-(-1)^{\frac{L_x}{2}}$ & $1$ & \m- & $(-1)^{\frac{N}{2}}$ & $1$ & $1$\\
\ms
o$\times$e & $\psi_0+\psi_{1/2}$ & $(-1)^{\frac{N}{2}}$ & $1$ & $(-1)^{\frac{L_y}{2}}$ & \m- & $(-1)^{\frac{N}{2}}$ & $1$ & $1$\\
& $\psi_0-\psi_{1/2}$ & $(-1)^{\frac{N}{2}}$ & $1$ & $-(-1)^{\frac{L_y}{2}}$ & \m- & $(-1)^{\frac{N}{2}}$ & $1$ & $1$\\
\br
\end{tabular}
\end{table}

\subsection{General remarks for rearrangement transformations}

Most of the transformations that we shall study below involve rearranging the spins on the lattice. An operator $\mathcal{O}$ describing such a transformation is unitary and acts as
\begin{equation}\label{trans}
\mathcal{O} \sigma^\alpha_j \mathcal{O}^\dag = \sigma^\alpha_{d^{-1}(j)},
\qquad j=\{1,2,\ldots,N\}, \qquad \alpha=x,y,z,
\end{equation}
where $\sigma^\alpha_j$ is the $\alpha$ Pauli operator acting on the spin sitting at the site with index $j$ and $d^{-1}(j)$ is a bijective map from $\{1,2,\ldots,N\}$ to $\{1,2,\ldots,N\}$ describing the rearrangement. Let us for notational convenience define
\begin{equation}
\psi_{a,b}(p(j),\zeta_j,s_j,\tau)=\psi_{a,b}(s_1,\ldots,s_N),
\end{equation}
where $p(j)$ specifies the choice of ordering of the labels as in \eref{psip}. The action of $\mathcal{O}$ on the state \eref{psiab} is
\begin{eqnarray}
\fl \mathcal{O}|\psi_{a,b}\rangle
=\mathcal{C}_{a,b}\mathcal{O}\sum_{s_1,\ldots,s_N} \psi_{a,b}(j,\zeta_j,s_j,\tau)|s_1,\ldots,s_N\rangle\nonumber\\
=\mathcal{C}_{a,b}\mathcal{O} \psi_{a,b}(j,\zeta_j,\sigma_j^z,\tau) \sum_{s_1,\ldots,s_N} |s_1,\ldots,s_N\rangle\nonumber\\
=\mathcal{C}_{a,b}\mathcal{O}\psi_{a,b}(j,\zeta_j,\sigma_j^z,\tau)\mathcal{O}^\dag \mathcal{O}\sum_{s_1,\ldots,s_N}|s_1,\ldots,s_N\rangle\nonumber\\
=\mathcal{C}_{a,b}\mathcal{O}\psi_{a,b}(j,\zeta_j,\sigma_j^z,\tau)\mathcal{O}^\dag \sum_{s_1,\ldots,s_N}|s_1,\ldots,s_N\rangle\nonumber\\
=\mathcal{C}_{a,b}\psi_{a,b}(j,\zeta_j,\sigma_{d^{-1}(j)}^z,\tau) \sum_{s_1,\ldots,s_N}|s_1,\ldots,s_N\rangle\nonumber\\
=\mathcal{C}_{a,b}\sum_{s_1,\ldots,s_N}\psi_{a,b}(j,\zeta_j,s_{d^{-1}(j)},\tau)|s_1,\ldots,s_N\rangle.
\end{eqnarray}
We can rewrite
\begin{equation}\label{rewrite}
\fl\psi_{a,b}(j,\zeta_j,s_{d^{-1}(j)},\tau)
=\psi_{a,b}(d(j),\zeta_{d(j)},s_j,\tau)
=(-1)^\mathcal{S}\psi_{a,b}(j,\zeta_{d(j)},s_j,\tau),
\end{equation}
where $\mathcal{S}$ is the number of times one has to swap two labels to change the numbering from $d(j)$ to $j$. The first equality in \eref{rewrite} follows from a relabelling of indices, and the second equality follows from \eref{swap}, which trivially generalizes to the case of twisted boundary conditions. To find the transformation properties of the wavefunction under $\mathcal{O}$, we thus only need to determine $\mathcal{S}$ and $\psi_{a,b}(j,\zeta_{d(j)},s_j,\tau)$. Since the delta function factor and the Marshall sign factor do not depend on $\zeta_j$, it is sufficient to consider the transformation properties of the centre of mass factor and the Jastrow factor. For later convenience, let us also define $\kappa_d$ through the relation
\begin{equation}\label{kappa}
\prod_{i<j}E(\zeta_{d(i)}-\zeta_{d(j)},\tau)^{(s_is_j+1)/2}
=\kappa_d\prod_{i<j} E(\zeta_i-\zeta_j,\tau)^{(s_i s_j+1)/2}.
\end{equation}

\subsection{Spin flip}\label{Sec:spinflip}

Let us first consider the spin flip operator $F=\prod_{j=1}^N\sigma_j^x$, which is not of the rearrangement type. It transforms $s_j$ into $-s_j$. The delta function and the Jastrow factor are easily seen to be invariant under this operation. The Marshall sign factor transforms as
\begin{eqnarray}\label{chims}
\fl\chi_{-\mathbf{s}}=\prod_{n=1}^N(-1)^{(n-1)(-s_n+1)/2}=\prod_{n=1}^N(-1)^{(n-1)(s_n-1)/2}
=\chi_\mathbf{s}\prod_{n=1}^N(-1)^{-(n-1)}\nonumber\\
=\chi_\mathbf{s}(-1)^{(N-1)N/2}=\chi_\mathbf{s}(-1)^{N/2},
\end{eqnarray}
and the transformation of the centre of mass factor follows from \eref{A4}. Altogether
\begin{equation}\label{F}
F\mathcal{C}^{-1}_{a,b}|\psi_{a,b}\rangle=(-1)^{N/2}\mathcal{C}^{-1}_{-a,-b}|\psi_{-a,-b}\rangle.
\end{equation}
Note that the factors $\mathcal{C}^{-1}_{a,b}$ and $\mathcal{C}^{-1}_{-a,-b}$ simply remove the normalization factors from $|\psi_{a,b}\rangle$ and $|\psi_{-a,-b}\rangle$, respectively. Some special cases of \eref{F} are
\begin{eqnarray}
F|\psi_{0,0}\rangle=(-1)^{N/2}|\psi_{0,0}\rangle,\\
F|\psi_{1/2,0}\rangle=(-1)^{N/2}\frac{\mathcal{C}_{1/2,0}}{\mathcal{C}_{-1/2,0}}|\psi_{-1/2,0}\rangle
=(-1)^{N/2}|\psi_{1/2,0}\rangle,\nonumber\\
F|\psi_{0,1/2}\rangle=(-1)^{N/2}\frac{\mathcal{C}_{0,1/2}}{\mathcal{C}_{0,-1/2}}|\psi_{0,-1/2}\rangle
=(-1)^{N/2}|\psi_{0,1/2}\rangle,\nonumber\\
F|\psi_{1/2,1/2}\rangle=(-1)^{N/2}\frac{\mathcal{C}_{1/2,1/2}}{\mathcal{C}_{-1/2,-1/2}}|\psi_{-1/2,-1/2}\rangle
=-(-1)^{N/2}|\psi_{1/2,1/2}\rangle,\nonumber
\end{eqnarray}
where we have used \eref{WFtwist}, \eref{psiab}, \eref{A2} and \eref{A3}. In particular, $\psi_{0,0}$ and $\psi_{1/2,0}$ are both eigenstates with eigenvalue $(-1)^{N/2}$ as stated in table \ref{Tab:eigen}.

\subsection{Translation in the $x$-direction}

\begin{figure}
\begin{indented}\item[]
\includegraphics[width=0.6\columnwidth]{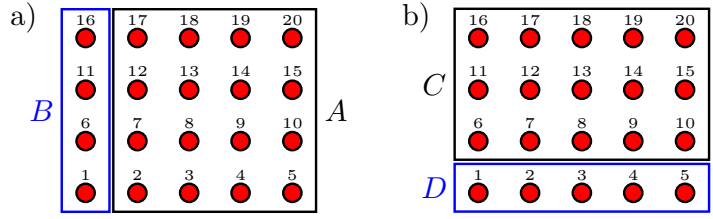}
\caption{The sets $A$, $B$, $C$, and $D$ for a $5\times4$ lattice. \label{Fig:sets}}
\end{indented}
\end{figure}

The action of the translation operator $T_x$ is to move the spins one lattice constant to the right, except for the rightmost column of spins which is wrapped around to the opposite edge. $T_x$ is of the rearrangement type and is defined through the map
\begin{equation}
d(j)=\left\{
\begin{array}{ll}
j-1 & {\rm for\ } j \in A, \\
j+(L_x-1) & {\rm for\ } j \in B,
\end{array}\right.
\end{equation}
which moves the lattice one lattice constant to the left. Here,
\begin{eqnarray}
A&=\{n+mL_x+1|n=1,\dots,L_x-1 \wedge m=0,1,\dots,L_y-1\},\\
B&=\{mL_x+1|m=0,\dots,L_y-1\},
\nonumber
\end{eqnarray}
i.e., $A$ is the set containing the indices of the sites in the $L_x-1$ rightmost columns of the lattice and $B$ is the set containing the indices of the sites in the leftmost column of the lattice as illustrated for a $5\times4$ lattice in figure \ref{Fig:sets}(a). This map has $\mathcal{S}=L_y(L_x-1)$.

To determine the transformation properties of the wavefunction, we note that
\begin{equation}
\zeta_{d(j)}=\left\{\begin{array}{ll}
\zeta_j-1/L_x & {\rm for\ } j \in A,\\
\zeta_j+(L_x-1)/L_x & {\rm for\ } j \in B.
\end{array}\right.
\end{equation}
Therefore
\begin{equation}
\fl\sum_{j=1}^N \zeta_{d(j)} s_j
=\sum_{j\in A}\left(\zeta_j-\frac{1}{L_x}\right)s_j
+\sum_{j\in B}\left(\zeta_j+\frac{L_x-1}{L_x}\right)s_j
=\sum_{j=1}^N\zeta_j s_j+\sum_{j\in B}s_j,
\end{equation}
and from \eref{A5} it then follows that
\begin{equation}
\thetaf{a}{b}{\sum_{j=1}^N \zeta_{d(j)} s_j}{2\tau}
=\thetaf{a}{b}{\sum_{j=1}^N \zeta_j s_j}{2\tau}
\rme^{2\pi\rmi a\sum_{j \in B} s_j}.
\end{equation}
Regarding the Jastrow factor, we observe that
\begin{equation}
\fl E(\zeta_{d(i)}-\zeta_{d(j)},\tau)
=\left\{\begin{array}{ll}
E(\zeta_i-\zeta_j,\tau) & {\rm for\ } (i\in A \wedge j\in A)
\vee (i\in B \wedge j\in B),\\
-E(\zeta_i-\zeta_j,\tau) & {\rm for\ } (i\in A \wedge j\in B)
\vee (i\in B \wedge j\in A).
\end{array}
\right.
\end{equation}
The case $i\in A$ and $j\in B$ follows from
\begin{eqnarray}
\fl E(\zeta_{d(i)}-\zeta_{d(j)},\tau)
=E\left(\zeta_i-\frac{1}{L_x}-\zeta_j-\frac{L_x-1}{L_x},\tau\right)
=E(\zeta_i-\zeta_j-1,\tau)\nonumber\\
=-E(\zeta_i-\zeta_j,\tau),
\end{eqnarray}
where we have used \eref{A12}, and the case $i\in B$ and $j\in A$ then follows from \eref{A11}. For $\kappa_d$ (see \eref{kappa}), we therefore get
\begin{equation}
\fl\kappa_d=(-1)^{\sum_{i\in A,j\in B}(s_is_j+1)/2}
=(-1)^{-\left(\sum_{j\in B}s_j\right)^2/2+L_y^2(L_x-1)/2}
=(-1)^{L_y^2(L_x-2)/2}.
\end{equation}
Collecting all the factors, we arrive at
\begin{equation}\label{Tx}
T_x|\psi_{a,b}\rangle=\left\{\begin{array}{ll}
(-1)^{L_x/2}\rme^{2\pi\rmi a\sum_{j\in B}\sigma^z_j}|\psi_{a,b}\rangle
& {\rm for\ }L_y{\rm\ odd},\\
\rme^{2\pi\rmi a\sum_{j\in B}\sigma^z_j}|\psi_{a,b}\rangle & {\rm for\ }L_y{\rm\ even}.
\end{array}\right.
\end{equation}
For $a\in\{0,1/2\}$, i.e., for $\theta_x=0$, we can replace $\sum_{j\in B}\sigma_j^z$ by $L_y$, and it follows that $\psi_{0,b}$ and $\psi_{1/2,b}$ are eigenstates of $T_x$ with the eigenvalues given in table \ref{Tab:eigen}.

\subsection{Translation in the $y$-direction}

The translation operator $T_y$ in the $y$-direction is described by the map
\begin{equation}
d(j)=\left\{
\begin{array}{ll}
j-L_x & {\rm for\ } j \in C, \\
j+(L_y-1)L_x & {\rm for\ } j \in D,
\end{array}\right.
\end{equation}
which moves the lattice one lattice constant in the negative $y$-direction. Here
\begin{eqnarray}\label{CD}
C&=\{L_x+1,L_x+2,\ldots,L_xL_y\},\\
D&=\{1,2,\dots,L_x\},
\nonumber
\end{eqnarray}
i.e., $C$ is the set of indices of the $L_y-1$ uppermost rows of the lattice and $D$ is the set of indices of the lowermost row of the lattice as illustrated for a $5\times4$ lattice in figure \ref{Fig:sets}(b). This map has $\mathcal{S}=L_x(L_y-1)$.

We note that
\begin{equation}
\zeta_{d(j)}=\left\{
\begin{array}{ll}
\zeta_j-\rmi/L_x & {\rm for\ } j \in C,\\
\zeta_j+\rmi(L_y-1)/L_x & {\rm for\ } j \in D.
\end{array}\right.
\end{equation}
Therefore
\begin{eqnarray}
\fl\sum_{j=1}^N\zeta_{d(j)} s_j=\sum_{j\in C}\left(\zeta_j
-\frac{\rmi}{L_x}\right)s_j+\sum_{j\in D}\left(\zeta_j
+\rmi\frac{L_y-1}{L_x}\right)s_j
=\sum_{j=1}^N\zeta_{j}s_{j}+\tau\sum_{j\in D}s_{j},
\end{eqnarray}
and, using \eref{A6} with $p\rightarrow p/2$ and $\tau\rightarrow2\tau$, it follows that
\begin{eqnarray}
\fl \thetaf{a}{b}{\sum_{j=1}^N \zeta_{d(j)} s_j}{2\tau}
=\thetaf{a+\frac{L_x}{2}}{b}{\sum_{j=1}^N\zeta_js_j}{2\tau}
\rme^{-\rmi\pi\tau(\sum_{j\in D}s_j)^2/2-\rmi\pi\sum_{j\in D}s_j (\sum_{j=1}^N\zeta_js_j+b)}.\nonumber\\
\end{eqnarray}
To derive the transformation of the Jastrow factor, we observe that
\begin{equation}
\fl E(\zeta_{d(i)}-\zeta_{d(j)},\tau)
=\left\{\begin{array}{ll}
E(\zeta_i-\zeta_j,\tau) & {\rm for\ } i\in C \wedge j\in C,\\
E(\zeta_i-\zeta_j,\tau) & {\rm for\ } i\in D \wedge j\in D,\\
\rme^{-\rmi\pi-\rmi\pi\tau-2\pi\rmi(\zeta_i-\zeta_j)}
E(\zeta_i-\zeta_j,\tau) & {\rm for\ } i\in D \wedge j\in C.
\end{array}
\right.
\end{equation}
The case $i\in C$ and $j\in D$ is not relevant, because this automatically implies $i>j$ according to \eref{CD}. The case $i\in D$ and $j\in C$ follows from
\begin{eqnarray}
\fl E(\zeta_{d(i)}-\zeta_{d(j)},\tau)
=E\left(\zeta_i+\rmi\frac{L_y-1}{L_x}-\zeta_j+\rmi\frac{1}{L_x},\tau\right)
=E(\zeta_i-\zeta_j+\tau,\tau)\nonumber\\
=\rme^{-\rmi\pi-\rmi\pi\tau-2\pi\rmi(\zeta_i-\zeta_j)}E(\zeta_i-\zeta_j,\tau),
\end{eqnarray}
where we have used \eref{A13}. $\kappa_d$ (see \eref{kappa}) is therefore
\begin{equation}
\kappa_d=(-1)^{\sum_{i\in D,j\in C}(s_is_j+1)/2}\rme^{-\rmi\pi\sum_{i\in D,j\in C}[\tau+2(\zeta_i-\zeta_j)](s_is_j+1)/2}.
\end{equation}
Note that
\begin{eqnarray}
\fl\sum_{i\in D,j\in C}[\tau+2(\zeta_i-\zeta_j)]
=\rmi\frac{L_y}{L_x}L_x^2(L_y-1)
+2L_x(L_y-1)\sum_{i\in D}\zeta_i-2L_x\sum_{j\in C} \zeta_j\nonumber\\
=\rmi L_yL_x(L_y-1)
+2L_xL_y\sum_{i\in D}\zeta_i\nonumber\\
=\rmi L_yL_x(L_y-1)
+2L_xL_y(-\rmi)\frac{L_y-1}{2L_x}L_x=0.
\end{eqnarray}
Therefore
\begin{eqnarray}
\fl\kappa_d=(-1)^{\sum_{i\in D,j\in C}(s_is_j+1)/2}\rme^{-\rmi\pi\sum_{i\in D,j\in C}[\tau+2(\zeta_i-\zeta_j)]s_is_j/2}\nonumber\\
=(-1)^{L_x^2(L_y-2)/2}\rme^{\rmi\pi\tau(\sum_{i\in D} s_i)^2/2+\rmi\pi\sum_{i=1}^N\zeta_is_i\sum_{j\in D}s_j}.
\end{eqnarray}
Collecting all the factors, we conclude that
\begin{equation}\label{Ty}
T_y\mathcal{C}^{-1}_{a,b}|\psi_{a,b}\rangle=\left\{\begin{array}{ll}
(-1)^{L_y/2}\rme^{-\rmi\pi b\sum_{j\in D}\sigma^z_j} \mathcal{C}^{-1}_{a+\frac{1}{2},b}|\psi_{a+\frac{1}{2},b}\rangle
& {\rm for\ }L_x{\rm\ odd},\\
\rme^{-\rmi\pi b\sum_{j\in D}\sigma^z_j} \mathcal{C}^{-1}_{a,b}|\psi_{a,b}\rangle
& {\rm for\ }L_x {\rm\ even}.
\end{array} \right.
\end{equation}
It follows that $T_y$ transforms states in the subspace spanned by $\psi_{a,0}$ and $\psi_{a+1/2,0}$ into states in the same subspace, and one can therefore easily diagonalize $T_y$ in these subspaces. The result for $a=0$ is given in table \ref{Tab:eigen}.

\subsection{Transformed translation operators}

As a side remark, we show that the action of the operators
\begin{equation}
\tilde{T_x}=UT_xU^\dag, \qquad \tilde{T_y}=UT_yU^\dag,
\end{equation}
on the wavefunctions, where $U$ is the unitary operator
\begin{equation}
U={\rm exp}\left[\frac{\rmi}{2}\sum_{n=0}^{L_x-1}\sum_{m=0}^{Ly-1}
\left(\frac{n\theta_x}{L_x}+\frac{m\theta_y}{L_y}\right)\sigma^z_{n+mL_x+1}\right],
\end{equation}
is in fact simpler than the action of $T_x$ and $T_y$. We first find explicit expressions for $\tilde{T}_x$ and $\tilde{T}_y$ by writing
$\tilde{T_d}=U(T_d U^\dag T^\dag_d)T_d$, $d=x,y$, and evaluating the expression in brackets using \eref{trans} with $\mathcal{O}=T_d$. This gives
\begin{eqnarray}
\tilde{T}_x={\rm exp}\left(-\frac{\rmi}{2}\theta_x\sum_{j\in B}\sigma^z_j
+\frac{\rmi}{2}\frac{\theta_x}{L_x}\sum_{j=1}^N\sigma^z_j\right)T_x,\\
\tilde{T}_y={\rm exp}\left(-\frac{\rmi}{2}\theta_y\sum_{j\in D}\sigma^z_j
+\frac{\rmi}{2}\frac{\theta_y}{L_y}\sum_{j=1}^N\sigma^z_j\right)T_y.
\end{eqnarray}
Combining this result with \eref{Tx} and \eref{Ty} and remembering \eref{ab}, we get
\begin{eqnarray}
\tilde{T}_x\mathcal{C}^{-1}_{a,b}|\psi_{a,b}\rangle=\left\{\begin{array}{ll}
(-1)^{2k}(-1)^{L_x/2}\mathcal{C}^{-1}_{a,b}|\psi_{a,b}\rangle
& {\rm for\ }L_y{\rm\ odd},\\
\mathcal{C}^{-1}_{a,b}|\psi_{a,b}\rangle & {\rm for\ }L_y{\rm\ even},
\end{array}\right.\label{tildeTx}\\
\tilde{T}_y\mathcal{C}^{-1}_{a,b}|\psi_{a,b}\rangle=\left\{\begin{array}{ll}
(-1)^{L_y/2}\mathcal{C}^{-1}_{a+1/2,b}|\psi_{a+\frac{1}{2},b}\rangle
& {\rm for\ }L_x{\rm\ odd},\\
\mathcal{C}^{-1}_{a,b}|\psi_{a,b}\rangle
& {\rm for\ }L_x {\rm\ even},
\end{array} \right.\label{tildeTy}
\end{eqnarray}
where $k=a-\theta_x/(4\pi)\in\{0,1/2\}$. We thus conclude that $\psi_{a,b}$ is an eigenstate of $\tilde{T}_x$ and $\tilde{T}_y$ for $L_x$ even, whereas $\psi_{a,b}\pm\psi_{a+1/2,b}$ are eigenstates for $L_x$ odd.

\subsection{Rotation by $90^{\circ}$}

In this subsection, we assume $L_x=L_y$ and study the action of a rotation by $90^\circ$. A rotation of the lattice by $-90^\circ$ is described by the map
\begin{equation}
d(n+mL_x+1)=m+(L_x-1-n)L_x+1,
\end{equation}
where $n=0,1,\ldots,L_x-1$ and $m=0,1,\ldots,L_x-1$. For this map $\mathcal{S}=3\times N/4$.

Let us note that
\begin{equation}
\zeta_{d(j)}=-\rmi\zeta_j=\zeta_j/\tau, \qquad \tau=\rmi=-1/\rmi=-1/\tau.
\end{equation}
We thus need to compute $\psi_{a,b}(j,\zeta_j/\tau,s_j,-1/\tau)$. The transformation done on this wavefunction is precisely the modular $S$-transformation (see, e.g., \cite{yellow}), which takes
\begin{equation}
\zeta_j\rightarrow\frac{\zeta_j}{\tau}, \qquad \tau\rightarrow-\frac{1}{\tau},
\end{equation}
and is defined for general $\tau$. We have
\begin{eqnarray}
\fl\thetaf{a}{b}{\frac{\zeta}{\tau}}{-\frac{2}{\tau}}
=\sum_{n\in\mathbb{Z}}\rme^{-\pi\rmi\frac{2}{\tau}(n+a)^2+2\pi \rmi(n+a)(\frac{\zeta}{\tau}+b)}
=\sum_{k\in\mathbb{Z}}\int_{-\infty}^\infty \rme^{2\pi\rmi kx-\pi \rmi\frac{2}{\tau}(x+a)^2+2\pi\rmi(x+a)(\frac{\zeta}{\tau}+b)}dx\nonumber\\
=\sqrt{\frac{\tau}{2\rmi}}\sum_{k\in\mathbb{Z}}\rme^{-\rmi2\pi ak
+\rmi\frac{\pi\tau}{2}(k+\frac{\zeta}{\tau}+b)^2}
=\sqrt{\frac{\tau}{2\rmi}}\sum_{\mu=0}^1\sum_{n\in\mathbb{Z}}\rme^{-\rmi2\pi a(2n+\mu)
+\rmi\frac{\pi\tau}{2}(2n+\mu+\frac{\zeta}{\tau}+b)^2}\nonumber\\
=\sqrt{\frac{\tau}{2\rmi}}\sum_{\mu=0}^1\sum_{n\in\mathbb{Z}}\rme^{
2\pi\rmi\tau(n+\frac{\mu}{2}+\frac{b}{2})^2
+2\pi\rmi(n+\frac{\mu}{2}+\frac{b}{2})(\zeta-2a)
+\frac{\pi\rmi\zeta^2}{2\tau}+2\pi\rmi ab}\nonumber\\
=\sqrt{\frac{\tau}{2\rmi}}
\rme^{\pi\rmi\zeta^2/(2\tau)+2\pi\rmi ab}
\sum_{\mu=0}^1 \thetaf{\frac{b}{2}+\frac{\mu}{2}}{-2a}{\zeta}{2\tau}.
\end{eqnarray}
A similar computation gives
\begin{equation}
\theta_1\left(\frac{\zeta}{\tau},-\frac{1}{\tau}\right)=
-\rmi(-\rmi\tau)^{1/2}\rme^{\rmi\pi \zeta^2/\tau} \theta_1(\zeta,\tau),
\end{equation}
which, together with \eref{Edef}, implies
\begin{equation}
E\left(\frac{\zeta_i-\zeta_j}{\tau},-\frac{1}{\tau}\right)=\tau^{-1}\rme^{\rmi\pi (\zeta_i-\zeta_j)^2/\tau}E(\zeta_i-\zeta_j,\tau).
\end{equation}
For the $L_x\times L_x$ lattice, we have $\sum_j\zeta_j=0$ and $\sum_j\zeta_j^2=0$, and therefore
\begin{equation}
\fl\prod_{i<j}\rme^{\rmi\pi (\zeta_i-\zeta_j)^2(s_is_j+1)/(2\tau)}=
\rme^{\rmi\pi \sum_i\sum_j(\zeta_i-\zeta_j)^2(s_is_j+1)/(4\tau)}
=\rme^{-\rmi\pi(\sum_{j=1}^N\zeta_js_j)^2/(2\tau)}.
\end{equation}
It follows that $\kappa_d=\tau^{-N(N-2)/4}\exp[-\rmi\pi(\sum_{j=1}^N\zeta_js_j)^2/(2\tau)]$. Collecting the factors and using the facts that $\tau=\rmi$ and that $N/4$ is an integer in our case, we find
\begin{equation}\label{R90}
R_{90^\circ}\mathcal{C}^{-1}_{a,b}|\psi_{a,b}\rangle=\frac{\rme^{2\pi\rmi ab}}{\sqrt{2}}
\left(\mathcal{C}^{-1}_{b/2,-2a}|\psi_{b/2,-2a}\rangle
+\mathcal{C}^{-1}_{b/2+1/2,-2a}|\psi_{b/2+1/2,-2a}\rangle\right).
\end{equation}
Particular cases are
\begin{eqnarray}
R_{90^\circ}\mathcal{C}^{-1}_{0,0}|\psi_{0,0}\rangle=\frac{1}{\sqrt{2}}
\left(\mathcal{C}^{-1}_{0,0}|\psi_{0,0}\rangle
+\mathcal{C}^{-1}_{1/2,0}|\psi_{1/2,0}\rangle\right),\\
R_{90^\circ}\mathcal{C}^{-1}_{1/2,0}|\psi_{1/2,0}\rangle=\frac{1}{\sqrt{2}}
\left(\mathcal{C}^{-1}_{0,0}|\psi_{0,0}\rangle
-\mathcal{C}^{-1}_{1/2,0}|\psi_{1/2,0}\rangle\right).\nonumber
\end{eqnarray}
From this one easily derives the results in table \ref{Tab:eigen}. Note also that the combination
\begin{equation}
|\psi_{0,0}(s_1,\ldots,s_N)|^2+|\psi_{1/2,0}(s_1,\ldots,s_N)|^2
\end{equation}
is invariant under the rotation by $90^\circ$. This is a consequence of the fact that the correlator \eref{corcb} of the complete field is modular invariant and that the absolute value of the factor $\eta(\tau)\prod_{i<j}E(\zeta_i-\zeta_j,\tau)^{1/2}$ that we removed from the conformal blocks in defining $\psi_{0,0}$ and $\psi_{1/2,0}$ in section \ref{Sec:singlet} is invariant under a $90^\circ$ rotation for the considered lattice. Simple eigenstates can also be found for $a\in\{1/4,3/4\}$ and $b=1/2$.

\subsection{Rotation by $180^{\circ}$}

The rotation by $180^\circ$ has
\begin{equation}
d(j)=N-j, \quad \zeta_{d(j)}=-\zeta_j, \quad \mathcal{S}=N/2.
\end{equation}
The transformation of the centre of mass factor follows from \eref{A8}, and the Jastrow factor gives the factor $\kappa_d=1$. Altogether, we therefore get
\begin{equation}\label{R180}
R_{180^\circ}\mathcal{C}^{-1}_{a,b}|\psi_{a,b}\rangle= (-1)^{N/2}\mathcal{C}^{-1}_{-a,-b}|\psi_{-a,-b}\rangle,
\end{equation}
which is the same result as for the spin flip operator.

\subsection{Time reversal combined with reflection}

In the FQH setting, time reversal corresponds to inverting the direction of the magnetic field. The action of the time reversal operator $\Theta$ on a general spin state
\begin{equation}
|\psi\rangle=\sum_{s_1,\ldots,s_N}\psi(s_1,\ldots,s_N)|s_1,\ldots,s_N\rangle
\end{equation}
is \cite{sakurai}
\begin{equation}
\Theta|\psi\rangle=
\rmi^N\sum_{s_1,\ldots,s_N}(-1)^{\sum_{j=1}^N(s_j+1/2)}\psi(-s_1,\ldots,-s_N)^*
|s_1,\ldots,s_N\rangle.
\end{equation}
Therefore time reversal takes $\mathcal{C}_{a,b}\psi_{a,b}(j,\zeta_j,s_j,\tau)$ into
$\mathcal{C}_{a,b}\psi^*_{a,b}(j,\zeta_j,-s_j,\tau)$, where we choose $\mathcal{C}_{a,b}$ to be real. The delta function factor is not affected by this transformation, and the Marshall sign factor is changed by the factor $(-1)^{N/2}$ as in \eref{chims}. Since $\tau$ is imaginary in our case, we get from \eref{A1} that
\begin{eqnarray}
\thetaf{a}{b}{-\sum_{j=1}^N\zeta_js_j}{2\tau}^*
=\thetaf{a}{-b}{\sum_{j=1}^N\zeta_j^*s_j}{2\tau},\\
E(\zeta_i-\zeta_j,\tau)^*=E(\zeta^*_i-\zeta^*_j,\tau).
\end{eqnarray}
The state is thus not invariant under time reversal, which is a generic feature for FQH wave functions.

We can get a second complex conjugation of $\zeta_j$ by acting with a reflection operator. The reflection operator $M_x$ in the $x$-direction has
\begin{equation}
d(n+mL_x+1)=L_x-n+mL_x, \quad \zeta_{d(j)}=-\zeta_j^*,
\quad \mathcal{S}=N/2,
\end{equation}
and the reflection operator $M_y$ in the $y$-direction has
\begin{equation}
d(n+mL_x+1)=n+(L_y-1-m)L_x+1, \quad \zeta_{d(j)}=\zeta_j^*,
\quad \mathcal{S}=N/2.
\end{equation}
Applying one of these operators and using \eref{A4} and \eref{A11}, we get
\begin{eqnarray}
M_x\Theta\mathcal{C}^{-1}_{a,b}|\psi_{a,b}\rangle=
\mathcal{C}^{-1}_{-a,b}|\psi_{-a,b}\rangle,\label{MxT}\\
M_y\Theta\mathcal{C}^{-1}_{a,b}|\psi_{a,b}\rangle=
\mathcal{C}^{-1}_{a,-b}|\psi_{a,-b}\rangle,\label{MyT}
\end{eqnarray}
which, together with \eref{A2}, leads to the results in table \ref{Tab:eigen}.

\section{Many-body Chern number and other topological properties}\label{Sec:chern}

\begin{figure}
\begin{indented}\item[]
\includegraphics[width=0.8\textwidth]{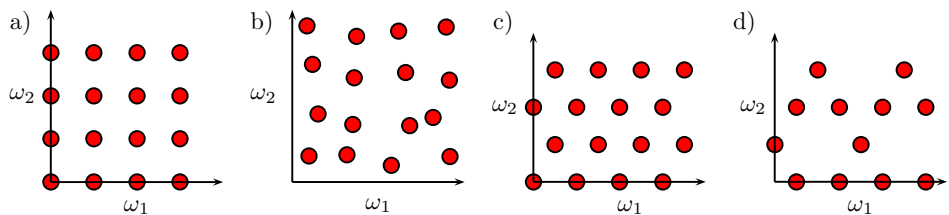}
\caption{The many-body Chern number of the two CFT states is one for the $4\times4$ square lattice in a), the randomly distorted $4\times4$ square lattice in b), the triangular lattice in c), and the kagome lattice in d). \label{Fig:chern}}
\end{indented}
\end{figure}

A main reason for studying FQH states is their nontrivial topological properties. For the KL states, the topological properties derive from the center of mass factor \cite{torus1,RZ96}. By now a number of measures have been identified that can be used to describe topological states, and some of these have already been used in \cite{nsc2} and \cite{nsc3} to demonstrate the topological nature of the CFT state on an irregular lattice on the sphere and the CFT states on square lattices on the torus. For the generalization of the KL states on the torus to arbitrary lattices presented in the present paper to be useful, it is important that the topological properties are preserved when the lattice is deformed away from the square lattice. To test this property, we compute the many-body Chern number for the lattices shown in figure \ref{Fig:chern}, and in all cases we get the expected value $1$. (The computation is done as described in detail in \cite{hatsugai,hafezi}, and we use the wavefunction \eref{WFtwist} with twisted boundary conditions.)

\begin{figure}
\begin{indented}\item[]
\includegraphics[width=0.55\textwidth]{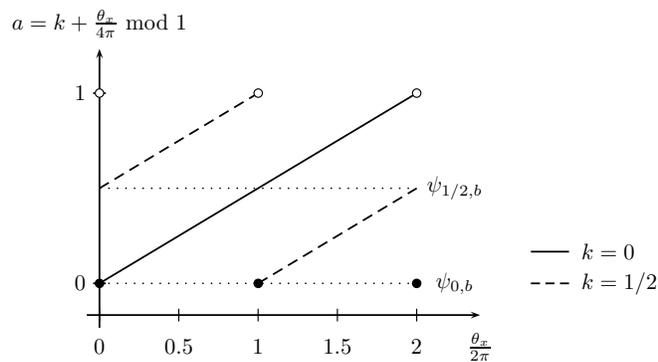}
\caption{When $\theta_x$ increases from $0$ to $2\pi$, $a$ increases by $1/2$. Since the wavefunction only depends on $a$ modulus $1$ (see \eref{A2}), it follows that $\psi_{0,b}$ is transformed into $\psi_{1/2,b}$ and vice versa. If $\theta_x$ is increased to $4\pi$, we are back to the starting point.}\label{Fig:twist}
\end{indented}
\end{figure}

The number of degenerate ground states on higher genus surfaces is an important quantity to partially characterize topological states. Degeneracies may, however, arise for other reasons than topology. To talk about topologically degenerate states, the local structure of the states must be the same, i.e., the expectation value of any local operator must be the same for all the states in the thermodynamic limit. Local indistinguishability has been demonstrated for the two states in \eref{WFT} for the case of a square lattice in \cite{nsc3}. One can observe from \eref{WFtwist} that the two states on the torus can be transformed into each other by changing the twist angles. Since the local structure is changed little by this operation when the number of spins is large, this is also a sign of topology. Specifically, if $\theta_x$ is changed from $0$ to $2\pi$, $a$ increases by $1/2$, and from \eref{A2} it then follows that the states are transformed as
\begin{eqnarray}
\psi_{0,0}(s_1,\ldots,s_N)\to \psi_{1/2,0}(s_1,\ldots,s_N),\\
\psi_{1/2,0}(s_1,\ldots,s_N)\to \psi_{0,0}(s_1,\ldots,s_N).
\end{eqnarray}
This is illustrated in figure \ref{Fig:twist}. If instead we increase $\theta_y$ from $0$ to $2\pi$, it follows from \eref{WFtwist} and \eref{A3} that the states are transformed as
\begin{eqnarray}
\fl\psi_{0,0}(s_1,\ldots,s_N)+\psi_{1/2,0}(s_1,\ldots,s_N)\to \psi_{0,0}(s_1,\ldots,s_N)-\psi_{1/2,0}(s_1,\ldots,s_N),\\
\fl\psi_{0,0}(s_1,\ldots,s_N)-\psi_{1/2,0}(s_1,\ldots,s_N)\to \psi_{0,0}(s_1,\ldots,s_N)+\psi_{1/2,0}(s_1,\ldots,s_N).
\end{eqnarray}

\section{Conclusion}\label{Sec:conclusion}

In conclusion, we have found that the KL states on the torus can be written as conformal blocks of primary fields from the $SU(2)_1$ WZW CFT. This representation of the states reveals an interesting underlying mathematical structure and gives a natural generalization of the KL states on the torus to arbitrary lattices. It also provides an easy way to ensure that the states are singlet states. In addition, we have found that the many-body Chern number for different lattices is unity and that the two states are transformed into each other under a twist of the boundary conditions. These findings suggest that the topological properties are preserved when the lattice is transformed away from the square lattice. Finally, we have analyzed the symmetry properties of the CFT states on the torus for the case of an $L_x\times L_y$ square lattice, which allowed us in certain cases to construct linear combinations of the states that are guaranteed to be orthogonal because they have different symmetries. Our work also shows that the KL states in different geometries can be obtained by evaluating the same conformal correlator in different geometries, and we believe that this holds more generally. The CFT representation found in the present paper constitutes an interesting starting point for further analytical investigations of the states. In \cite{tu2}, CFT representations of lattice Laughlin states with general filling factor $1/q$, where $q$ is an integer, have been found on the Riemann sphere, and using these results, we note that the present work could be straightforwardly generalized to obtain lattice Laughlin states with filling factor $1/q$ on the torus. Higher genus versions of the CFT states fulfilling the CFT fusion rules can also be obtained. Another interesting perspective is to make similar constructions for other FQH states.

\ack
The authors acknowledge discussions with J. Ignacio Cirac, Denis Bernard, and Hong-Hao Tu. This work has been partially funded by EU through SIQS grant (FP7 600645) and by FIS2012-33642, QUITEMAD (CAM), and the Severo Ochoa Program.

\appendix

\section{Properties of the Riemann theta function}\label{Sec:AppA}

The Riemann theta function is defined as
\begin{equation}\label{A1}
\thetaf{a}{b}{\zeta}{\tau}=
\sum_{n \in \mathbb{Z}}\rme^{\rmi\pi\tau(n+a)^2+2\pi\rmi(n+a) (\zeta+b)},
\end{equation}
where $a$ and $b$ are real numbers, $\zeta$ is complex, and $\tau$ is complex with $\mathrm{Im}(\tau)>0$. From this definition, one easily derives the following identities
\begin{eqnarray}
\thetaf{a+c}{b}{\zeta}{\tau}=\thetaf{a}{b}{\zeta}{\tau},
\quad c\in\mathbb{Z},\label{A2}\\
\thetaf{a}{b+c}{\zeta}{\tau}
=\rme^{2\pi \rmi ac}\thetaf{a}{b}{\zeta}{\tau},
\quad c\in\mathbb{Z},\label{A3}\\
\thetaf{a}{b}{-\zeta}{\tau}=\thetaf{-a}{-b}{\zeta}{\tau},\label{A4}\\
\thetaf{a}{b}{\zeta+c}{\tau}
=\rme^{2\pi \rmi ac}\thetaf{a}{b}{\zeta}{\tau},
\quad c\in\mathbb{Z},\label{A5}\\
\thetaf{a}{b}{\zeta+p\tau}{\tau} = \rme^{-\rmi\pi\tau p^2-\rmi 2\pi p(\zeta+b)}
\thetaf{a+p}{b}{\zeta}{\tau}, \quad p\in\mathbb{R}.\label{A6}
\end{eqnarray}
Particular cases that we shall need are
\begin{eqnarray}
\thetaf{k}{0}{-\zeta}{\tau}=\thetaf{k}{0}{\zeta}{\tau}, \quad k=0,\frac{1}{2}, \label{A8} \\
\thetaf{k}{0}{\zeta\pm 1}{2\tau}=\rme^{\pm 2\pi\rmi k}\thetaf{k}{0}{\zeta}{2\tau}, \quad k=0,\frac{1}{2}, \label{A9} \\
\thetaf{k}{0}{\zeta \pm \tau}{2\tau}=\rme^{-\rmi\pi\tau/2\mp\pi \rmi \zeta}
\thetaf{1/2-k}{0}{\zeta}{2\tau}, \quad k=0,\frac{1}{2}, \label{A10}
\end{eqnarray}
and
\begin{eqnarray}
E(-\zeta,\tau)=-E(\zeta,\tau),\label{A11}\\
E(\zeta\pm1,\tau)=\rme^{\pm\rmi\pi} E(\zeta,\tau),\label{A12}\\
E(\zeta\pm\tau,\tau)=\rme^{-\rmi\pi\tau\mp\rmi\pi\mp2\pi\rmi \zeta} E(\zeta,\tau),\label{A13}
\end{eqnarray}
where $E(\zeta,\tau)$ is the prime form defined in \eref{Edef}.

\section{Singlet property for four spins from Fay's trisecant identity}\label{Sec:AppB}

Consider the case of four spins. From \eref{WFT}, the wavefunctions on the torus read
\begin{eqnarray}
\fl|\psi_k\rangle=\mathcal{C}_k(\psi_{1,k}|+1,-1,+1,-1\rangle
+\psi_{2,k}|+1,-1,-1,+1\rangle+\psi_{3,k}|+1,+1,-1,-1\rangle\\
+\psi_{1,k}|-1,+1,-1,+1\rangle+\psi_{2,k}|-1,+1,+1,-1\rangle
+\psi_{3,k}|-1,-1,+1,+1\rangle),\nonumber
\end{eqnarray}
where
\begin{eqnarray}
\psi_{1,k}=E(\zeta_1-\zeta_3,\tau)E(\zeta_2-\zeta_4,\tau) \thetaf{k}{0}{\zeta_1-\zeta_2+\zeta_3-\zeta_4}{2\tau},
\label{36}\\
\psi_{2,k}=-E(\zeta_1-\zeta_4,\tau)E(\zeta_2-\zeta_3,\tau) \thetaf{k}{0}{\zeta_1-\zeta_2-\zeta_3+\zeta_4}{2\tau}, \nonumber\\
\psi_{3,k}=-E(\zeta_1-\zeta_2,\tau)E(\zeta_3-\zeta_4,\tau) \thetaf{k}{0}{\zeta_1+\zeta_2-\zeta_3-\zeta_4}{2\tau}. \nonumber
\end{eqnarray}
This state is a singlet if and only if
\begin{equation}\label{35}
\psi_{1,k}+\psi_{2,k}+\psi_{3,k}=0.
\end{equation}
Now we use (A3) in \cite{RZ96} to express the theta function with argument $2\tau$ as the product of two theta functions with argument $\tau$,
\begin{equation}
\thetaf{k}{0}{\zeta}{2\tau}=\frac{\eta(2\tau)}{\eta^2(\tau)}
\thetaf{k}{0}{\frac{\zeta}{2}+\frac{1}{4}}{\tau}
\thetaf{k}{0}{\frac{\zeta}{2}-\frac{1}{4}}{\tau}, \;\, k=0,\frac{1}{2},
\end{equation}
where $\eta$ is the Dedekind eta function defined in \eref{dedekind}. We can therefore write \eref{36} as
\begin{eqnarray}
\fl\psi_{1,k}=[\eta(2\tau)/\eta^2(\tau)]
E(\zeta_1-\zeta_3,\tau)E(\zeta_2-\zeta_4,\tau)\\
\times \thetaf{k}{0}{\frac{\zeta_1-\zeta_2+\zeta_3-\zeta_4}{2}+\frac{1}{4}}{\tau}
\thetaf{k}{0}{\frac{\zeta_1-\zeta_2+\zeta_3-\zeta_4}{2}-\frac{1}{4}}{\tau},
\nonumber\\
\fl\psi_{2,k}=-[\eta(2\tau)/\eta^2(\tau)]
E(\zeta_1-\zeta_4,\tau)E(\zeta_2-\zeta_3,\tau)\nonumber\\
\times \thetaf{k}{0}{\frac{\zeta_1-\zeta_2-\zeta_3+\zeta_4}{2}+\frac{1}{4}}{\tau}
\thetaf{k}{0}{\frac{\zeta_1-\zeta_2-\zeta_3+\zeta_4}{2}-\frac{1}{4}}{\tau},\nonumber\\
\fl\psi_{3,k}=-[\eta(2\tau)/\eta^2(\tau)]
E(\zeta_1-\zeta_2,\tau)E(\zeta_3-\zeta_4,\tau)\nonumber\\
\times \thetaf{k}{0}{\frac{\zeta_1+\zeta_2-\zeta_3-\zeta_4}{2}+\frac{1}{4}}{\tau}
\thetaf{k}{0}{\frac{\zeta_1+\zeta_2-\zeta_3-\zeta_4}{2}-\frac{1}{4}}{\tau}.\nonumber
\end{eqnarray}
These quantities satisfy \eref{35} as a consequence of Fay's trisecant identity \cite{F73,M83}
\begin{eqnarray}
\fl E(\zeta_1-\zeta_3,\tau)E(\zeta_2-\zeta_4,\tau)\thetaf{a}{b}{\zeta}{\tau} \thetaf{a}{b}{\zeta+\zeta_1+\zeta_3-\zeta_2-\zeta_4}{\tau}=\nonumber\\
E(\zeta_1-\zeta_4,\tau)E(\zeta_2-\zeta_3,\tau)\thetaf{a}{b}{\zeta+\zeta_1-\zeta_2}{\tau} \thetaf{a}{b}{\zeta+\zeta_3-\zeta_4}{\tau}\nonumber\\
+E(\zeta_1-\zeta_2,\tau)E(\zeta_3-\zeta_4,\tau)\thetaf{a}{b}{\zeta+\zeta_1-\zeta_4}{\tau} \thetaf{a}{b}{\zeta+\zeta_3-\zeta_2}{\tau}. \label{39}
\end{eqnarray}
This identity has already been used in CFT in connection with the bosonization formulas on the torus \cite{EE87}. To prove \eref{35} we have to choose $\zeta$ in \eref{39} as
\begin{equation}
\zeta=-\frac{\zeta_1-\zeta_2+\zeta_3-\zeta_4}{2}+\frac{1}{4}
\end{equation}
and use \eref{A8}.

\end{document}